\documentclass{aa}

\newcommand{\kms}{km\ s$^{-1}$}

\newcommand{\micron}{$\mu$m}
\newcommand{\HTWOO}{$\textrm{H}_2 \textrm{O}$}

\newcommand{\COiso}{$^{13}\textrm{CO}$}
\newcommand{\COmain}{$^{12}\textrm{CO}$}
\newcommand{\methane}{$\textrm{CH}_4$}
\newcommand{\ammonia}{$\textrm{NH}_3$}

\usepackage[caption=false]{subfig}
\usepackage{hyperref}
\usepackage{graphicx}
\usepackage{multirow}
\usepackage{txfonts}
\usepackage{color}

\begin{document} 

   \title{The $^{12}$CO/$^{13}$CO isotopologue ratio of a young, isolated brown dwarf}

   \subtitle{Possibly distinct formation pathways of super-Jupiters and brown dwarfs}
   \titlerunning{CO isotopologue ratio}
    
    \author{Yapeng Zhang \inst{1}, Ignas A.G. Snellen \inst{1}, Paul Molli\`ere \inst{2}}
        
   \institute{1: Leiden Observatory, Leiden University, Postbus 9513, 2300 RA, Leiden, The Netherlands \\
   2: Max-Planck-Institut f\"{u}r Astronomie , K\"{o}nigstuhl 17, 69117 Heidelberg, Germany   \\
   }
    \authorrunning{}
    
  \date{Received June 09, 2021; accepted Sep 21, 2021}

\abstract{Linking atmospheric characteristics of planets to their formation pathways is a central theme in the study of extrasolar planets. Although the $^{12}$C/$^{13}$C isotope ratio shows little variation in the solar system, the atmosphere of a super-Jupiter was recently shown to be rich in $^{13}$CO, possibly as a result of dominant ice accretion beyond the CO snowline during its formation. Carbon isotope ratios are therefore suggested to be a potential tracer of  formation pathways of planets.}
{In this paper we aim to measure the \COmain/\COiso\, isotopologue ratio of a young, isolated brown dwarf. While the general atmospheric characteristics of young, low mass brown dwarfs are expected to be very similar to those of super-Jupiters, their formation pathways may be different, leading to distinct isotopologue ratios. In addition, such objects allow high-dispersion spectroscopy at high signal-to-noise.}
{We analyse archival K-band spectra of the L dwarf 2MASS J03552337+1133437 taken with NIRSPEC at the Keck telescope. A free retrieval analysis is applied to the data using the radiative transfer code \texttt{petitRADTRANS} coupled to the nested sampling tool \texttt{PyMultiNest},  to determine the isotopologue ratio \COmain/\COiso\, in its atmosphere.}
{The isotopologue \COiso\, is detected in the atmosphere through the cross-correlation method at a signal-to-noise of $\sim$8.4. The detection significance is determined to be $\sim9.5\sigma$ using Bayesian model comparison between two retrieval models (including or excluding \COiso). We retrieve an isotopologue ratio \COmain/\COiso\, of $97^{+25}_{-18}$ (90\% uncertainty), marginally higher than the local interstellar standard. Its C/O ratio of $\sim$0.56 is consistent with the solar value.}
{Although only one super-Jupiter and one brown dwarf have now a measured \COmain/\COiso\, ratio, it is intriguing that they are different, possibly hinting to distinct formation pathways.
Albeit spectroscopic similarities, isolated brown dwarfs may experience a top-down formation via gravitational collapse, which resembles star formation, while giant exoplanets favourably form through core accretion, which potentially alters isotopologue ratios in their atmospheres depending on the material they accrete from protoplanetary disks. This further emphasises atmospheric carbon isotopologue ratio as a tracer for the formation history of exoplanets. In the future, analyses such as presented here should be conducted on a wide range of exoplanets using medium to high-resolution spectroscopy to further assess planet formation processes.} 


   \maketitle
%

\section{Introduction}
Planet formation and evolution are expected to leave imprints on the observed spectra of exoplanets. In bridging the gap between the spectral characterization and formation mechanisms, isotopologue ratios have been suggested to be informative tracers of planet formation conditions and evolutionary history \citep{Clayton2004, Zhang2021}. 
In the Solar System, deuterium/hydrogen (D/H) ratios demonstrate significant variations across planets, comets, and meteorites \citep{Altwegg2015}. While the D/H ratios in Jupiter and Saturn are consistent with the proto-solar value, Uranus and Neptune are found to be enhanced in deuterium \citep{Feuchtgruber2013}, attributed to a likely increased contribution from accretion of D-rich ices beyond the water snowline. The terrestrial planets Earth, Mars, and Venus have higher D/H ratios \citep{Drake2005}, indicating not only solid accretion, but also atmospheric loss. 
Therefore, isotope ratios in planetary atmospheres can reflect the material reservoir of the birth environment, the formation mechanism (via core-accretion or gravitational collapse), the relative importance of gas/ice accretion, and the atmospheric evolution. 

In contrast to D/H ratios, carbon isotope ratios are found to be roughly constant in the Solar System \citep{Woods2009}, and therefore less diagnostic. However, the recent measurement of the isotopologue abundance ratio \COmain/\COiso\, in an exoplanet may require a reassessment of its diagnostic value. The \COiso\, isotopologue was detected at a significance of 6$\sigma$ in the atmosphere of the super-Jupiter TYC 8998 b. Intriguingly, the atmosphere is reported to be \COiso-rich with a \COmain/\COiso\, abundance ratio of 31, which is significantly lower than the local interstellar standard at $\sim$68 \citep{Zhang2021}.  A formation outside the CO snowline with a large contribution from ice-accretion (as an analogy to the D-enrichment in solar-system planets) has been invoked to explain the enrichment. Since the solar-system planets are thought to be formed within the CO snowline, no substantial enrichment in $^{13}$C is expected, because the bulk of their carbon reservoirs originate from CO gas in the inner disk, where $^{13}$C is not enhanced. 
It therefore suggests that the atmospheric carbon isotopologue ratios could also shed light upon the formation history of exoplanets, especially for the directly imaged population which are observed at wide orbits beyond the CO snowline. 

Several carbon fractionation processes in molecular clouds, young stellar objects (YSOs), and protoplanetary disks are suggested to alter the CO isotopologue ratios in the birth environment of planets, including isotopic ion exchange reactions \citep{Langer1984, Milam2005}, isotope-selective photodissociation \citep{Bally1982, vanDishoeck1988, Visser2009, Miotello2014}, and ice/gas isotopologue partitioning \citep{Acharyya2007, Smith2015}. Depending on the location of the proto-planets and the material (gas/ice) they accrete from the environment, the isotopologue ratios in the atmospheres may deviate from the ISM standard. 
Therefore, detailed modeling of carbon fractionation in protoplanetary disks coupled with planet formation models have the potential to locate the birthplace and identify the formation mechanism of planets, like suggested for carbon-to-oxygen (C/O) ratios \citep{Oberg2011, Madhusudhan2014, Mordasini2016, Cridland2020}. Moreover, combining evidence from different observables, such as $^{12}$C/$^{13}$C, D/H, and C/O ratios, allows for a more comprehensive understanding of the formation process. 

Compared to the D/H ratios ($\sim$10$^{-5}$) which potentially can be probed via HDO/\HTWOO\, and CH$_3$D/\methane\, \citep{Morley2019, Molliere:Snellen:2019}, the $^{12}$C/$^{13}$C ratios ($\lesssim$100) are more readily detectable and attained from the ground with high-resolution spectroscopy targeting \COmain/\COiso\, \citep{Molliere:Snellen:2019}.
In addition to the recent result for a super-Jupiter, carbon isotopologue ratios have been measured toward various sources beyond the Solar System, including the interstellar medium (ISM) \citep{Langer1993, Milam2005, Yan2019}, YSOs and protostars in gas and ices \citep{Boogert2000, Boogert2002, Pontoppidan2005, Jorgensen2016, Jorgensen2018, Smith2015}, giant stars, solar-type stars, M dwarfs  \citep{Sneden1986, Botelho2020, Tsuji2016, Crossfield2019}, but not yet towards brown dwarfs. 
In this paper, we carry out a similar analysis as \citet{Zhang2021} on archival high-resolution Keck/NIRSPEC data of a young, isolated brown dwarf. While its general atmospheric characteristics is expected to be very similar to that of super-Jupiters, its formation pathways may be different, leading to distinct isotopologue ratios. In addition, its spectrum can be studied at high signal-to-noise, and hence is more accessible.

We present the observations and spectrum extraction in Section~\ref{sec:method}, followed by a description of the retrieval model in Section~\ref{sec:retrieval}. The retrieval results and the measurement of CO isotopologue ratio can be found in Section~\ref{sec:result}, and are discussed in Section~\ref{sec:discussion}.
   
\section{Observations and Spectrum Extraction}\label{sec:method}

\begin{table}
\caption{Properties of 2M0355, including those derived in this work.} 
\label{tab:info}
\begin{tabular}{lcc}
\hline
\hline
\textbf{Parameter} & \textbf{Symbol} & \textbf{Value} \\ 
\hline
R.A. (J2000)$^a$ & $\alpha$ & 03:55:23.377 \\
Dec. (J2000)$^a$ & $\delta$ & +11:33:43.7 \\
Distance (pc)$^a$ & $d$ & $9.1 \pm 0.1$ \\
Systemic velocity $\textrm{(\kms)}^b$   & $v_\mathrm{sys}$ 		& $11.8 \pm 0.5$ \\
Rotational velocity $\textrm{(\kms)}^b$   & $v\sin (i) ^ *$ & $14.7 \pm 2.3$ \\
Spectral type$^c$ & & L5$\gamma$/L3\\
2MASS K-magnitude (mag)$^d$ & $K_\mathrm{mag}$ & $11.526 \pm 0.021$ \\
Age (Myr)$^e$ &  & $\sim125$ \\ 
Effective temperature $(\textrm{K})^e$ 	& $T_{\textrm{eff}}$ 	& $1430 \pm 40$ \\
Mass $(M_\mathrm{Jup})^f$ 		   			& $M_p$ 				& $19^{+7}_{-5}$ \\
Surface gravity $\textrm{(cgs)}^f$ 		& $\log g$ 				& $4.69\pm 0.15$ \\
Carbon-to-Oxygen ratio$^f$ & C/O & $0.56 \pm 0.02$ \\
CO isotopologue ratio$^f$ & \COmain/\COiso & $97^{+25}_{-18}$ \\
\hline
\end{tabular}
\footnotesize{$^*$ our work indicates a smaller rotation velocity than found in the literature.}
\tablebib{ $(a)$ \citet{GaiaDR2}; $(b)$ \citet{Blake2010, Bryan2018}; $(c)$ \citet{Cruz2009}; $(d)$ \citet{Cutri2003}; $(e)$ \citet{Faherty2013, Liu2013, Aller2016} $(f)$ this work.}

\end{table}

  \begin{figure*}
  \centering
  \includegraphics[width=\hsize]{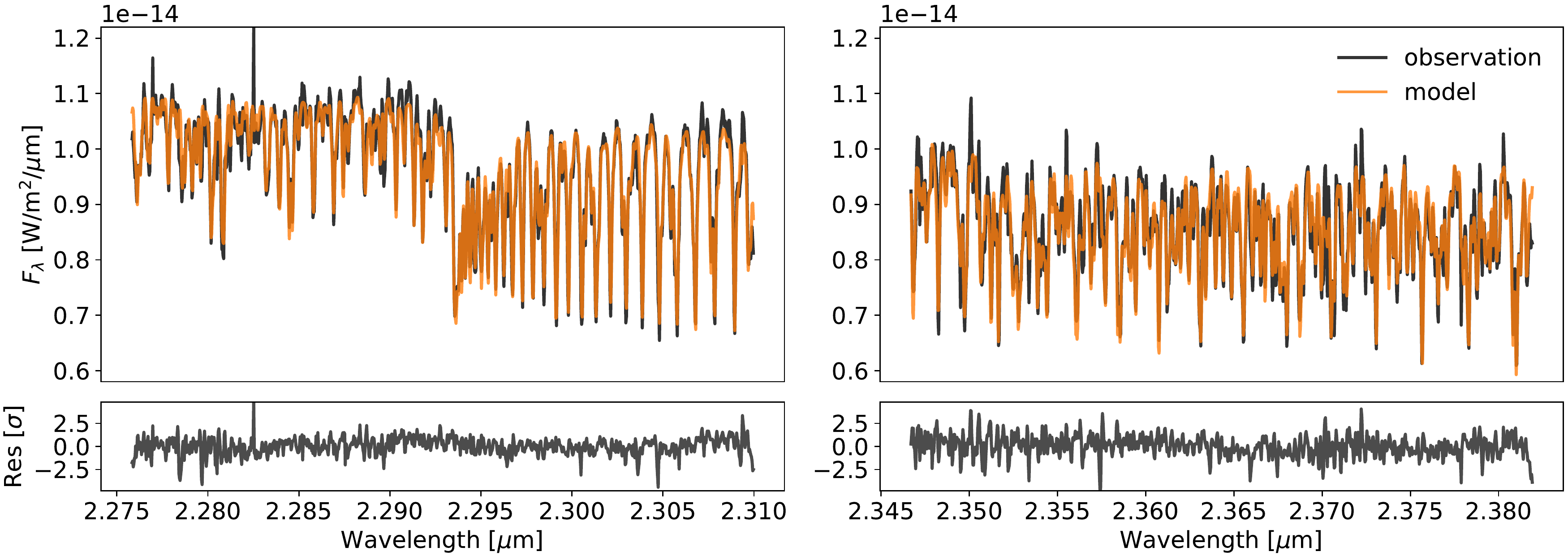}
      \caption{The last two orders (containing CO opacity) of the K-band spectrum of the brown dwarf 2M0355 taken with Keck/NIRSPEC. The orange line is the best-fit model with log($g$)=4.69, [Fe/H]=0.2, C/O=0.55, and \COmain/\COiso=97, obtained from the maximum likelihood model in the retrieval analysis. The bottom panel shows the residuals of the observed spectrum with respect to the model.
              }
         \label{fig:spec}
  \end{figure*}
  
2MASS J03552337+1133437 (hereafter 2M0355) is a free-floating young brown dwarf discovered from 2MASS \citep{Reid2008}, and is likely a member of AB Doradus moving group \citep{Liu2013}. With a distance of 9.1 parsec and a spectral type of L5$\gamma$ (with $\gamma$ denoting a very low surface gravity), it is among the nearest and reddest L dwarfs \citep{Cruz2009, Faherty2013}. The properties of 2M0355 are summarised in Table \ref{tab:info}. Its spectrum demonstrates signatures of low surface gravity and resembles those of directly imaged planetary mass objects, while being orders of magnitudes brighter and not contaminated by starlight. It is therefore an excellent target for high-resolution spectroscopic studies, providing insights into atmospheric properties under similar conditions as in exoplanets.

We used the archival K-band (2.03-2.38 \micron) spectra of 2M0355 taken with the near-infrared spectrograph NIRSPEC ($\lambda/\Delta\lambda \sim25\,000$) at the Keck II 10 m telescope \citep{McLean1998} on Janurary 13, 2017. The data were obtained in natural seeing mode with a 0.432x24\arcsec slit. The observations were performed with a standard ABBA nod pattern, resulting in 14 science exposures between 400 and 600 seconds, which amount to a total exposure time of two hours. The data have been previously used to measure the spin of the brown dwarf by \citet{Bryan2018}.

As for the pre-processing, we first corrected the data using dark and flat field calibration frames, and then differenced each nodded AB pair to subtract the sky background. This led to 2D differenced images with two spectral traces (positive and negative) for each order. In the subsequent analysis, We focussed on the last two orders with the wavelength coverage of 2.27 - 2.31 \micron\ (blue order) and 2.34 - 2.38 \micron\ (red order) respectively.
Subsequently, we straightened the 2D images on an order-by-order basis to align the curved spectral traces along the dispersion (x-)axis by fitting the traces with a third order polynomial. Similarly, we also adjust the spectral traces  to correct for the tilted spectral lines on the y-axis due to instrument geometry. The tilt was measured by fitting the brightest sky line in the flat-fielded raw image with a linear function, which was then applied to the differenced frames. We then combined all the rectified 2D images to a single frame for each order. 

From the combined 2D frames, we extracted 1D spectra for both the positive and negative trace, using the optimal extraction algorithm by \citet{Horne1986}. This method takes the weighted sum of the flux along the spatial dimension (the cross-dispersion axis), based on the empirical point spread function (PSF) constructed from the 2D frame, while rejecting outliers caused by cosmic rays or bad pixels. 
After obtaining 1D spectra for both traces, we add them up to form the master spectrum, resulting in an average SNR of $\sim$35 per unit of dispersion (pixel). This SNR value was estimated by comparing the master spectrum to a model spectrum obtained by the retrieval analysis (detailed in Section~\ref{sec:retrieval}). Using the statistical flux uncertainties as calculated in the spectrum extraction process (accounting for shot noise and readout noise), we expect the SNR to be $\sim$115, which differs from the former estimation by a factor of 3.3. This indicates the presence of correlated noise in the data and/or the imperfect modeling of telluric transmission and planetary spectra. Nevertheless, the SNR estimated via the comparison to models represents a lower limit to the data quality. We therefore base our further analysis on the assumption of this lower SNR.

To determine the wavelength solution for each order, we took advantage of the standard star spectra obtained under the same instrument configuration immediately before the observation, assuming that the wavelength solution remains the same. We compared the standard star spectrum with a telluric transmission model generated with ESO sky model calculator \texttt{SkyCalc}\footnote{\url{https://www.eso.org/observing/etc/skycalc}} \citep{Noll2012, Jones2013} and fitted the wavelengths as a function of pixel positions using $\sim$25 telluric lines spread across each order with a third order polynomial. This solution was then applied to the target spectrum.

We then used the ESO sky software tool \texttt{Molecfit} v3.0.1 \citep{Smette2015} to perform telluric corrections on the master spectrum. The tool uses a Line-By-Line Radiative Transfer Model (LBLRTM) to derive telluric atmospheric transmission spectra that can be fitted to observations. For model fitting, we selected two wavelength regions, 2.276 - 2.293 \micron, and 2.347 - 2.382 \micron, which include strong telluric lines caused by \methane\, and \HTWOO. The software accounted for molecular abundances, instrument resolution, continuum level and wavelength solution that can fit observations best. Using the best-fit Gaussian kernel width, we estimated the spectral resolution to be $\lambda/\Delta\lambda \sim27\,500$. The atmospheric transmission model for the entire wavelength range was then derived based on the best-fit parameters. The model was removed from the master spectrum to obtain the telluric-corrected spectrum. Subsequently, we shifted the spectrum to the target's rest frame by accounting for the systemic and barycentric radial velocity. The flux was finally scaled to the K-band photometry of $1.05\times10^{-14}$ W/m$^2$/\micron\ \citep{Cutri2003}. The scaling factor was determined by generating the planetary model spectrum at a larger wavelength coverage, then integrating the flux over the band pass of the K-band filter, and comparing it to the photometric value.

We note that measuring the CO isotopologue ratio requires careful calibration of the broad-band spectral shape. Because the CO absorption features span across the two spectral orders in the observations, the CO abundance inferred from the spectrum is sensitive to slight changes in the relative flux between the two orders. To calibrate this, we inspected spectra of two standard stars taken immediately before and after the target observations. We reduced the standard star spectra following the same procedure as described above, and compared the telluric-corrected spectrum of the standard stars to a \texttt{PHOENIX} stellar model \citep{Husser2013} with an effective temperature of 8200 K as estimated for the stars \citep{GaiaDR2}. Since the standard star spectra are firmly in the Rayleigh-Jeans regime, potential uncertainties in effective temperature have negligible effects on the spectral shape. The comparison suggested that the observed flux ratio between the red and blue order was on average 3\% higher than that of the model.
We therefore decreased the flux in the red order by 3\% to compensate for this.

The final spectrum of 2M0355 is shown in Fig.~\ref{fig:spec}. 
The \COmain\, $\nu$=2-0 bandhead is clearly visible at 2.2935 \micron, while the first bandhead of \COiso\,at 2.3448 \micron\, falls outside of the detector. On the order of $\sim$30 \COiso\, lines are covered by the red order (see Fig.~\ref{fig:ccf} left panel).

\section{Retrieval Analysis}\label{sec:retrieval}

\subsection{Atmospheric retrieval model}

For our atmospheric free retrieval, in which all fitted parameters are varied freely under the condition of chemical equilibrium but  the temperature structure being unconstrained, we used a Bayesian framework composed of the radiative transfer tool \texttt{petitRADTRANS} (\texttt{pRT}) \citep{Molliere2019} and the nested sampling tool \texttt{PyMultiNest} \citep{Buchner2014}, which is a Python wrapper of the \texttt{MultiNest} method \citep{Feroz2009}. Synthetic emission spectra are generated by \texttt{pRT} using a set of inputs, including the temperature structure, chemical abundances, and surface gravity. The \texttt{PyMultiNest} samples the parameter space and derives the posterior distribution of the fit.

The forward modelling consists of three major components: the temperature model, the chemistry model, and the cloud model.
We parameterise the temperature-pressure (T-P) profile using four temperature knots spaced evenly on a log scale pressure within 0.02 to 5 bar, where the contribution function of the observed spectrum peaks (see upper right panel in Fig.~\ref{fig:corner}). The entire T-P profile is obtained by spline interpolation of the four temperature knots in the log space of pressure. The temperature profile outside this range is barely probed by the observations, and is simply considered to be isothermal in our model. There is no physical reasoning behind this T-P profile, therefore imposing few prior constraints on the solution.

The chemistry model used in our retrievals assumes chemical equilibrium as detailed in \citet{Molliere2017,Molliere2020}. In short, the chemical abundances are determined via interpolation in a chemical equilibrium table using pressure P, temperature T, carbon-to-oxygen ratio C/O, and metallicity [Fe/H] as inputs. 

As for the cloud model, the same setup as in \citet{Molliere2020} is implemented, using the \citet{Ackerman2001} model. It introduces additional 4 free parameters: the mass fraction of the cloud species at the cloud base $X_0^c$, the settling parameter $f_\mathrm{sed}$ (controlling the thickness of the cloud above the cloud base), the vertical eddy diffusion coefficient $K_\mathrm{zz}$ (effectively determining the particle size), and the width of the log-normal particle size distribution $\sigma_g$. The location of the cloud base $P_\mathrm{{base}}$ is determined by intersecting the condensation curve of the cloud species with the T-P profile of the atmosphere.
MgSiO$_3$ and Fe clouds are expected as the dominant cloud species in L dwarfs \citep{Morley2012}. Here we only consider the MgSiO$_3$ clouds, because Fe is not expected to be the dominant aerosol composition according to the microphysics model by \citet{Gao2020}, and (even if the cloud forms) it condensates at a higher temperature which occurs at lower altitudes than the photosphere of the target.

\subsection{Retrieving 2M0355}
The cloudy retrieval model has 14 free parameters: $R_p$, log($g$), [Fe/H], C/O, \COmain/\COiso, $X_0^\mathrm{MgSiO_3}$, $f_\mathrm{sed}$, $K_\mathrm{zz}$, $\sigma_g$, $v$sin($i$) and four temperature knots for the T-P profile. The priors of these free parameters are listed in Table~\ref{tab:prior}. 
We included \HTWOO, \methane, \ammonia, \COmain, and the isotopologue \COiso\ as line opacity species. The model also accounts for the Rayleigh scattering of H$_2$, He, the collision induced absorption of H$_2$-H$_2$, H$_2$-He, the scattering and absorption cross sections of crystalline, irregularly shaped MgSiO$_3$ cloud particles.

We used the line-by-line mode of \texttt{pRT} to calculate the emission spectra at high spectral resolution. To speed up the calculation, we took every fifth point of opacity tables with $\lambda/\Delta\lambda \sim10^6$. This sampling factor of 5 has been benchmarked against the original sampling to ensure unbiased inference of parameters. The synthetic high-resolution spectra were rotationally broadened by $v$sin($i$) and convolved with a Gaussian kernel to match the resolving power of the instrument ($\lambda/\Delta\lambda \sim27\,500$), then binned to the wavelength grid of the observed spectrum (2020 data points in total), and scaled to the observed flux according to $R_p$ and distance of the target.

The retrievals were performed by \texttt{PyMultiNest} in Importance Nested Sampling mode with a constant efficiency of 5\%. It uses 2000 live points to sample the parameter space and derives the posterior abundances. 

\begin{table}
\caption{Priors and posteriors (90\% uncertainties) of the 2M0355 retrievals. 
} 
\label{tab:prior}
\begin{tabular}{llll}
\hline
\hline
\textbf{Parameter} & \textbf{Prior} & \textbf{Posterior} & \textbf{Posterior2} \\ 
\hline
log(\COmain/\COiso) & $\mathcal{U}$(-12,\ 0) & $-1.99 \pm 0.10$ & $-2.03 \pm 0.09$ \\
$R_p$ [$R_\mathrm{Jup}$] & $\mathcal{U}$(0.3,\ 2.5) & $0.97 \pm 0.02$ & $1.13 \pm 0.03$\\
log($g$) [cgs] & $\mathcal{U}$(3.0,\ 6.0) & $4.69 \pm 0.15$   & $4.32 \pm 0.15$ \\
$\rm [Fe/H]$ & $\mathcal{U}$(-1.5,\ 1.5) & $0.21 \pm 0.10$ & $0.05 \pm 0.11$  \\
C/O & $\mathcal{U}$(0.1,\ 1.5) & $0.56 \pm 0.02$ & $0.55 \pm 0.02$\\
T$_0$ [K] & $\mathcal{U}$(1500,\ 4000) & $2322 \pm 138$ & $2906 \pm 224$ \\
T$_1$ [K] & $\mathcal{U}$(0.5,\ 1)*T$_0$ & $1731 \pm 17$ & $1671 \pm 35$ \\
T$_2$ [K] & $\mathcal{U}$(0.5,\ 1)*T$_1$ & $1640 \pm 18$ & $1478 \pm 20$ \\
T$_3$ [K] & $\mathcal{U}$(0.5,\ 1)*T$_2$ & $1452 \pm 21$ & $1446 \pm 16$ \\
log($\Tilde{X}_0^\mathrm{MgSiO_3}$)& $\mathcal{U}$(-2.3,\ 1) & $-1.3 \pm 1.2$ & -\\
$f_\mathrm{sed}$ & $\mathcal{U}$(0,\ 10) & $5.8 \pm 4.1$ & $8.0 \pm 2.2$\\ 
log($K_\mathrm{zz}$) & $\mathcal{U}$(5,\ 13) & $10.5 \pm 3.2$ & $10.0 \pm 1.8$ \\
$\sigma_g$ & $\mathcal{U}$(1.05,\ 3) & $2.0 \pm 0.8$ & $1.89 \pm 0.82$\\
$v$sin($i$) & $\mathcal{U}$(0,\ 20) & $2.0 \pm 1.3$ & $1.2 \pm 1.3$\\
log($\tau_\mathrm{cloud}$) & $\mathcal{U}$(0,\ 1) & - & $ 0.02 \pm 0.03$\\
\hline
\end{tabular}
\tablefoot{The column Posterior shows values for the nominal model, while Posterior2 refers to the alternative model with enforced clouds. 
$\mathcal{U}$(a, b) represents a uniform distribution ranging from a to b. The $\Tilde{X}_0^\mathrm{MgSiO_3}$ denotes $X_0^\mathrm{MgSiO_3}$/$X_\mathrm{eq}^\mathrm{MgSiO_3}$, where $X_\mathrm{eq}^\mathrm{MgSiO_3}$ is  the  mass  fraction  predicted  for  the cloud  species  when  assuming  equilibrium  condensation  at the cloud base location.}
\end{table}

\section{Results}\label{sec:result}

\subsection{Retrieval results}
\begin{figure*}
  \centering
  \includegraphics[width=\hsize]{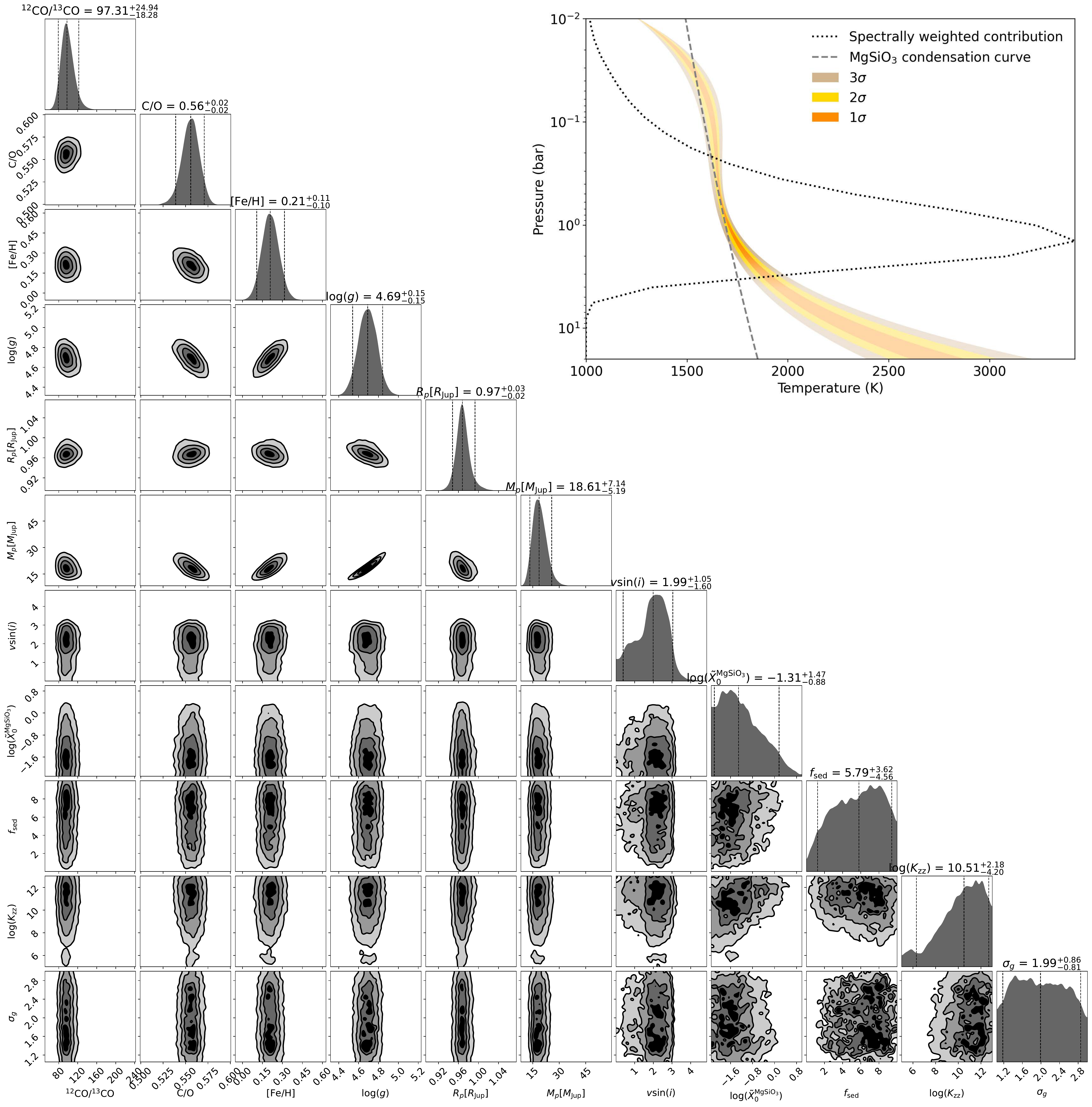}
      \caption{Retrieved parameters and temperature structure of 2M0355 using the nominal model. Upper right panel: retrieved temperature-pressure confidence envelopes. The coloured regions represent 1$\sigma$, 2$\sigma$, 3$\sigma$ confidence envelopes. The black dotted line shows the flux average of the emission contribution function. The gray dashed line represents the condensation curve of MgSiO$_3$ clouds.
      Lower left panel: posterior distributions of parameters. The vertical dashed lines denote the 5\%, 50\%, 95\% quantiles (90\% uncertainties) of the distribution. 
              }
         \label{fig:corner}
  \end{figure*}

The retrieval results are shown in Fig.\ref{fig:corner}, including the retrieved temperature-pressure profile and the posterior distribution of the free parameters. The central values of the inferred parameters and their 90\% uncertainties are summarised in Table~\ref{tab:prior}, including \COmain/\COiso\,$=97^{+25}_{-18}$, C/O $=0.56^{+0.02}_{-0.02}$, [Fe/H] $=0.21^{+0.11}_{-0.10}$, log($g$) $=4.69^{+0.15}_{-0.15}$, and $R_p$ $=0.97^{+0.03}_{-0.02}$. 
It suggests that 2M0355 has a solar C/O ratio and a super-solar metallicity. The inferred mass of $19^{+7}_{-5}$ $M_\mathrm{Jup}$ is in line with previous estimations using evolutionary models based on its photometry and membership (age) of AB Dor moving group \citep{Faherty2013}. Using the
evolutionary models for low-mass objects by \citet{Baraffe2015}, we estimate the radius of a $\sim20M_\mathrm{Jup}$ dwarf to be $\sim1.15R_\mathrm{Jup}$ for an age of 120 Myr.
The radius constrained from our retrieval model ($\sim0.97 R_\mathrm{Jup}$) is smaller than that expected from evolutionary tracks. This is likely associated with the temperature structure in the retrieval models. We note that the alternative retrieval model (described below) converges to different T-P profiles, leading to radii in agreement with the estimation from evolutionary models (see Table~\ref{tab:prior}).

Moreover, we retrieved a much smaller rotational velocity (< 4 \kms) than the value (14.7 \kms) found by \citet{Bryan2018} using the same data, as listed in Table~\ref{tab:info}. Broadening synthetic spectra with a $v$sin($i$) of 14.7 \kms\, result in too broad spectral lines to fit the observation with a $\lambda/\Delta\lambda \sim27\,500$. In our case, the widths of spectral lines are therefore dominated by the instrument resolution instead of the spin of the target. Regardless, the rotational broadening shows no impact on the inference of the isotopologue ratio.

It shows in Fig.\ref{fig:corner} that the properties of clouds are barely constrained. The inferred values of the parameters are similar as in the cloud-free case that we tested, indicating that the retrieval converges to solutions with optically thin clouds. We suggest that this behaviour is a cooperative consequence of the freely adjustable temperature structure and properties of clouds. Since the presence of thick clouds can block the emission from high-temperature regions of the atmosphere below, the cloudy atmosphere is spectrally equivalent to the case of a cloudless atmosphere combined with a shallow (isothermal) temperature gradient. This effect has been put forward as an alternative (cloud-free) explanation for the red spectra of brown dwarfs and self-luminous exoplanets \citep{Tremblin2015, Tremblin2016, Tremblin2017, Tremblin2019}. Both cases provide good fits to the observed spectrum, but here the free retrieval model tends to converge to cloudless solutions, potentially because they occupy a larger volume in the parameter space than the cloudy solutions. The problem of too isothermal temperature structures in free retrievals of cloudy objects has also been discussed in \citet{Burningham2017, Molliere2020, Burningham2021}. Classical 1-d self-consistent calculations will predict a steeper temperature profile, requiring clouds to fit the data. 
It is therefore important to examine how the inference of other parameters is affected in both cases.

To obtain a more reasonable temperature structure incorporating the effect of clouds, we have to enforce the atmosphere to be cloudy with a free parameter $\tau_\mathrm{cloud}$, which artificially sets the optical depth of the cloud at the photosphere of the clear component of the atmosphere (where the major emission contribution lies assuming no clouds). In this case, we no longer fit for the cloud mass fraction $X_0^\mathrm{c}$. Instead, the optical depth of clouds at all pressure levels is scaled to fulfill the prescribed $\tau_\mathrm{cloud}$. A strict prior is set to enforce optically thick clouds ($\tau_\mathrm{cloud}$>1). Moreover, we also ensure that the prescribed $\tau_\mathrm{cloud}$ is physically plausible by constraining the scaling factor of the cloud optical depth to be less than $2(f_\mathrm{sed}+1)$. This factor $f_\mathrm{sed}+1$ results from the requirement that the cloud surface density (i.e. integrating the product of density with cloud mass fraction vertically) is constrained by the available mass above the cloud base before condensation takes place (i.e. with the equilibrium mass fraction $X_\mathrm{eq}^\mathrm{c}$). With the cloud mass fraction depending on the altitude via $X^\mathrm{c}(P) \propto P^{f_\mathrm{sed}}$, this leads to the constraint for the mass fraction at the cloud base: $X_0^\mathrm{c} \leq X_\mathrm{eq}^\mathrm{c} (f_\mathrm{sed}+1)$. We add a factor of two to lend some flexibility.

\begin{figure*}
  \centering
  \includegraphics[width=\hsize]{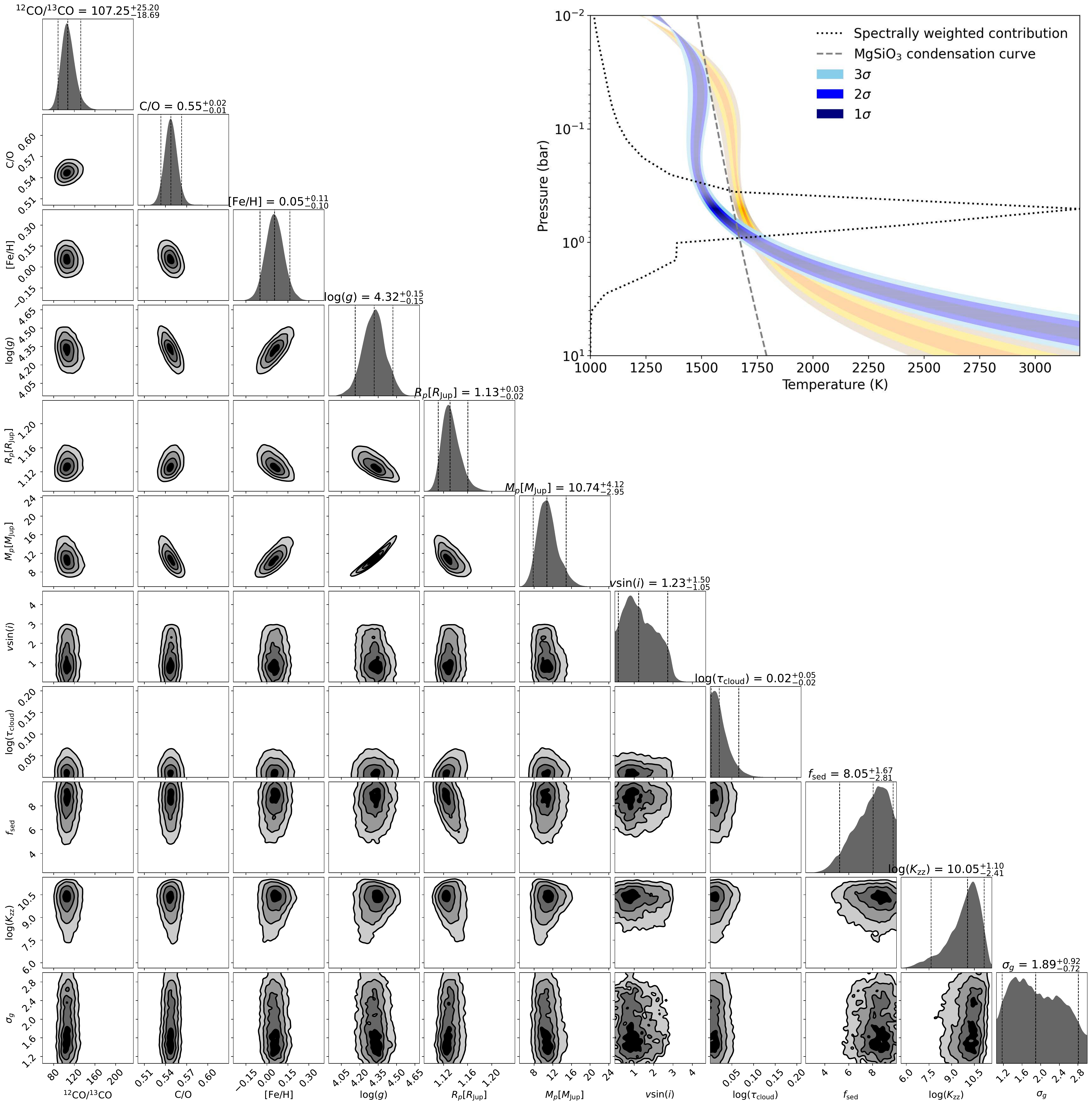}
      \caption{Similar as Fig.~\ref{fig:corner} but for the alternative model with enforced clouds. Upper right panel compares the temperature-pressure profiles of the nominal model in Fig.~\ref{fig:corner} (in orange) to this alternative model (in blue).}
         \label{fig:corner_cloudy}
  \end{figure*}

This alternative model results in more pronounced clouds and less isothermal T-P profiles as shown in Fig.~\ref{fig:corner_cloudy}. Comparing the retrieved values of other atmospheric properties (see Table~\ref{tab:prior}), we note that the atmospheric gravity and radius are significantly modified as a result of the heavy cloud and the altered temperature structure. The alternative model no longer probes as deep as the cloudless model because the clouds hide the region below. In addition, the metallicity becomes smaller when clouds are enforced in the retrieval, because the steeper temperature gradient requires less absorbing gas to
achieve a similar absorption in emergent flux, as also explained in \citet{Burningham2021}. Retrievals over a larger wavelength coverage (in particular, probing silicate absorption features at 10 $\mu$m) will reveal the nature of clouds and help better constrain other properties of the atmospheres \citep{Burningham2021}. Nevertheless, it is reassuring that the retrieval of C/O and \COmain/\COiso\, ratio does not strongly rely on the accurate inference of temperature structure and clouds, which was also found in \citet{Molliere2020, Burningham2021}. 

\subsection{\texorpdfstring{$^{13}$CO}{} detection}
\begin{figure*}
  \centering
  \includegraphics[width=\hsize]{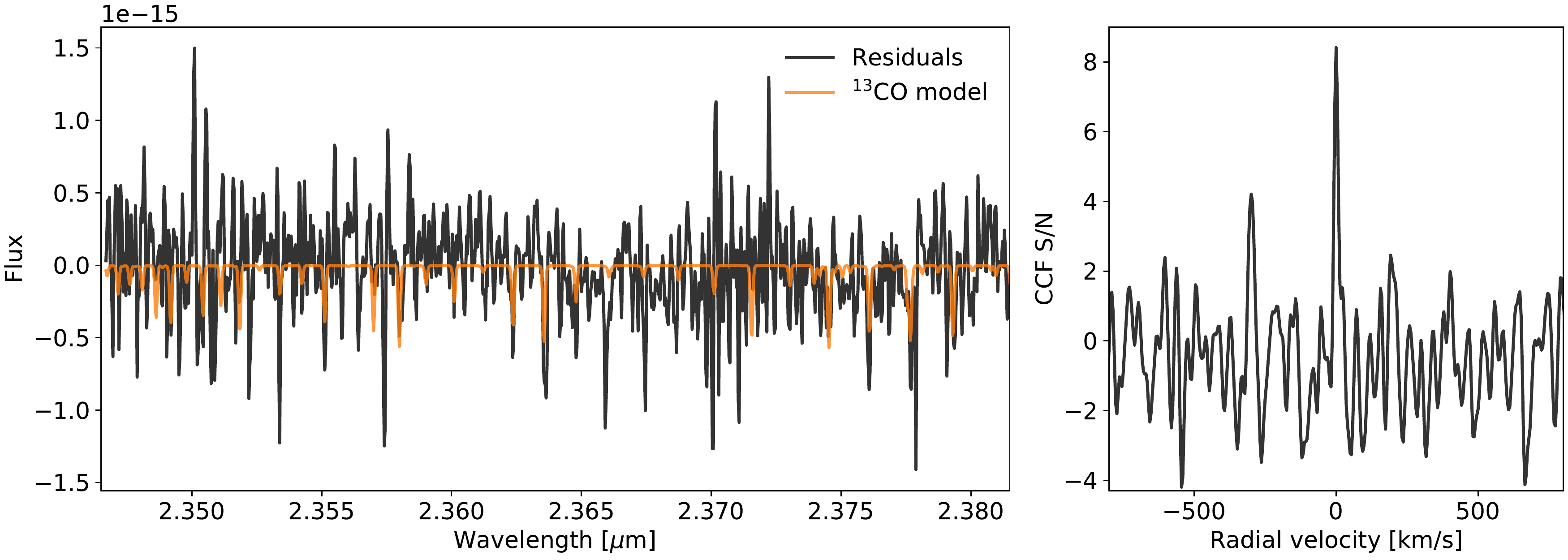}
      \caption{Left panel: observational residuals (observed spectrum minus best-fit model with the \COiso\, abundance set to zero) in black and \COiso\, model spectrum given the best-fit \COmain/\COiso\, ratio ($\sim$97) in orange. Right panel: cross-correlation function (CCF) between the \COiso\, model and residuals. The CCF is normalised by its standard deviation within the velocity ranges [-2200, -800] and [800, 2200] \kms, so that the y-axis represents the signal-to-noise of the CCF peak.
              }
         \label{fig:ccf}
  \end{figure*}

To reveal the presence of \COiso\, signal, we compare the observed spectrum to the retrieved best-fit model (log($g$)=4.69, [Fe/H]=0.2, C/O=0.55, and \COmain/\COiso=97) in Fig.~\ref{fig:spec}, and calculate the cross-correlation function (CCF) between the residuals (the observed spectrum minus the best-fit model without \COiso) and a \COiso\, model (Fig.~\ref{fig:ccf}). The \COiso\, model spectrum is constructed by taking the difference between the best-fit model and the same model with the \COiso\, abundance set to zero. As is shown in Fig.~\ref{fig:ccf} right panel, the CCF peaks at zero radial velocity, signifying the detection of \COiso\, at a signal-to-noise (S/N) ratio of 8.4. This value varies slightly depending on the choice of velocity ranges where the standard deviation of the noise is measured.

We also determine the significance of the \COiso\, detection using Bayesian model selection. We perform the retrieval again while excluding \COiso\, opacity and the parameter \COmain/\COiso. Comparing its Bayesian evidence ($Z$) to that of the full model (as listed in Table~\ref{tab:lnz}) allows us to assess the extent to which the model including \COiso\, is favoured by the observations. In Bayesian model comparison, the Bayes factor $B_m$ (calculated by the ratio of $Z$) is used as a proxy for the posterior odds ratio between two models. We then translate $B_m$ to the frequentist measures of significance following \citet{Benneke2013}. As a result, the difference of $\ln(Z)$ between two models is $\Delta\ln(Z)= \ln(B_m)= 44$, meaning that the observation favours the full model (including \COiso) at a significance level of $9.5\sigma$. The detection remains the same significance ($9.5\sigma$) when we use the alternative model with enhanced clouds.

\begin{table}
\caption{Results of Bayesian model comparison for the two retrieval models with and without \COiso.  
} 
\label{tab:lnz}
\begin{tabular}{cccc}
\hline
\hline
\textbf{Model} & \textbf{$\ln(Z)$} & \textbf{$\chi^2_\mathrm{best-fit}$} & $\ln(B_m)$ \\ 
\hline
Full model & -1181 & 2372 & 44 ($\sim9.5\sigma$) \\
\COiso\,excluded & -1225 & 2470 & -\\
\hline
\end{tabular}
\tablefoot{$Z$ denotes the Bayesian evidence, $\chi^2_\mathrm{best-fit}$ the best-fit $\chi^2$, and $B_m$ the Bayes factor.}
\end{table}

\section{Discussion}\label{sec:discussion}

\subsection{CO isotopologue ratio in 2M0355}

We found the CO isotopologue ratio in 2M0355 (\COmain/\COiso $=97^{+25}_{-18}$) to be similar to the carbon isotope ratio measured in for the Sun \citep[$93.5 \pm 3$  by][]{Lyons2018}, while marginally higher than the ratio in today's local ISM \citep[$68\pm 15$ by][]{Milam2005}, which is expected to be inherited by young objects. However, recent measurements of carbon isotope ratios towards a sample of solar-twins suggest a current local value of $\sim$81.3 \citep{Botelho2020}, which is within the uncertainty of our measurement. Therefore, it remains unclear whether the CO isotopologue ratio in 2M0355 is indeed higher than its local environment.

We note that high \COmain/\COiso\, ratios were also observed toward some diffuse clouds, molecular clouds and YSOs in the solar neighbourhood \citep{Lambert1994, Federman2003, Goto2003, Smith2015}.
If the high \COmain/\COiso\, ratios in molecular clouds and YSOs do not result from observational biases, it may not be surprising that the brown dwarf 2M0355 might inherit the high ratio from a \COiso-diminished parent cloud.
One possible explanation for the high ratios towards diffuse clouds is the isotope-selective photodissociation \citep{Bally1982, vanDishoeck1988, Visser2009}. Because of the lower column density of the rarer isotopologue, the self-shielding of \COiso\, kicks in at a deeper layer into the cloud. Consequently, \COiso\, is preferably destructed by far ultraviolet (FUV) radiation, increasing the \COmain/\COiso\, ratios in the ISM. However, this explanation is not as plausible for molecular clouds and YSOs, because these objects with high extinction are supposed to be well shielded from FUV radiation. The isotope-selective photodissociation is unlikely to significantly alter the globally isotopologue ratios \citep{vanDishoeck1988}. Other hypotheses were also suggested, such as higher excitation temperature in \COmain\, than in \COiso\, resulted from photon trapping in \COmain\, rotational transitions \citep{Goto2003}, CO gas/ice reservoir partitioning \citep{Smith2015}, and ISM enrichment caused by ejecta from carbon-rich giant stars or supernovae \citep{Crossfield2019}, but no general and conclusive interpretation.

\subsection{Implications for planet formation}

It is intriguing to compare the \COmain/\COiso\, measurements between the L dwarf 2M0355 ($\sim$97) and the wide-orbit exoplanet TYC 8998 b \citep[$\sim$31 by][]{Zhang2021}. Despite similarities in several aspects, the young super-Jupiter has possibly undergone a different formation pathway from the isolated brown dwarf, leading to the difference in isotopologue ratio in their atmospheres. While isolated brown dwarfs may experience a top-down formation via gravitational collapse (which resembles star formation) or disk instability \citep{Boss1997, Kratter2016}, super-Jupiters possibly form through the bottom-up core-accretion (of planetesimals and pebbles) \citep{Pollack1996, Lambrechts2012}. Previous demographics on giant planets and brown dwarfs \citep{Nielsen2019, Bowler2020} also support the two distinct formation pathways.
As argued in \citet{Zhang2021}, the core-accretion scenario could possibly lead to \COiso\, enrichment through ice accretion, lowering the \COmain/\COiso\, ratio in observed super-Jupiter atmospheres.

Admittedly, it is preliminary to draw any conclusive interpretation based on a sample of two objects. Another caveat is that the two measurements towards the super-Jupiter and the L dwarf are carried out under different spectral resolution. It is so far unclear whether a smaller spectral resolving power can lead to any observational biases because it probes unresolved molecular lines. Hence, this calls for more future observations that enable homogeneous comparisons between different objects, to better understand the potential and limitations of carbon isotopologue ratio as a tracer for planet formation. It is also essential to interpret the isotopologue measurements in light of modeling of carbon fractionation in protoplanetary disks \citep{Miotello2014}, disk evolution \citep{Trapman2021}, CO ice chemistry \citep{Krijt2020}, and planet formation \citep{Cridland2020}, in order to provide constraints on the birth location, the relative importance of ice accretion, and possible vertical gas accretion processes during formation.

\section{Conclusion}\label{sec:conclusion}

We analysed archival high-resolution spectra (from 2.275 to 2.385 \micron) from Keck/NIRSPEC of the isolated brown dwarf 2M0335, and carried out a free retrieval analysis. 
After removing the retrieved best-fit model that includes only the main isotopologues, we detect a \COiso\, signal using the cross-correlation method with a S/N of $\sim$8.4. The detection significance is determined to be $\sim9.5\sigma$ with Bayesian model comparison between the two retrieval models including or excluding \COiso. 
The isotopologue ratio \COmain/\COiso\, is inferred to be $97^{+25}_{-18}$, similar to the value found in the Sun, while marginally higher than the local ISM standard. 
If this deviation is real, it is possibly inherited from its parent cloud that has a high \COmain/\COiso\, ratio somehow. 

Although based on only two objects, it is also intriguing to note the difference between carbon isotopologue ratios in giant exoplanets and brown dwarfs, potentially implying distinct formation pathways. Albeit spectroscopic similarities, brown dwarfs may experience a top-down formation via gravitational collapse, which resembles star formation, while giant exoplanets likely form through a bottom-up accretion process, which may alter isotopologue ratios in their atmospheres with dependencies on the accretion location and pathway in protoplanetary disks. This further emphasises the atmospheric carbon isotopologue ratio as a tracer for planet formation history. 
In the future, such analysis can be implemented on spectra of more exoplanets to measure the carbon isotopologue ratios in their atmospheres and shed light on the location and mechanisms of planet formation.

\begin{acknowledgements}
We thank the referee and editor for helpful comments. This research has made use of the Keck Observatory Archive (KOA), which is operated by the W. M. Keck Observatory and the NASA Exoplanet Science Institute (NExScI), under contract with the National Aeronautics and Space Administration. This work was performed using the compute resources from the Academic Leiden Interdisciplinary Cluster Environment (ALICE) provided by Leiden University.
Y.Z. and I.S. acknowledge funding from the European Research Council (ERC) under the European Union's Horizon 2020 research and innovation program under grant agreement No. 694513.
P.M. acknowledges support from the European Research Council under the European Union’s Horizon 2020 research and innovation program under grant agreement No. 832428. 

\end{acknowledgements}

\bibliographystyle{aa} 
\bibliography{ref} 

\end{document}